\documentclass[sigconf, breaklinks=true]{acmart}
\newcommand{\an}[2]{#1} % Switch to #2 to anonymize

\usepackage{balance}
\usepackage[inline]{enumitem}
\usepackage{microtype}
\usepackage{multirow}

\copyrightyear{2024}
\acmYear{2024}
\setcopyright{rightsretained}
\acmConference[SIGCSE 2024]{Proceedings of the 55th ACM Technical Symposium on Computer Science Education V. 1}{March 20--23, 2024}{Portland, OR, USA}
\acmBooktitle{Proceedings of the 55th ACM Technical Symposium on Computer Science Education V. 1 (SIGCSE 2024), March 20--23, 2024, Portland, OR, USA}
\acmDOI{10.1145/3626252.3630754}
\acmISBN{979-8-4007-0423-9/24/03}

\makeatletter
\gdef\@copyrightpermission{
  \begin{minipage}{0.3\columnwidth}
   \href{https://creativecommons.org/licenses/by/4.0/}{\includegraphics[width=0.90\textwidth]{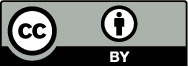}}
  \end{minipage}\hfill
  \begin{minipage}{0.7\columnwidth}
   \href{https://creativecommons.org/licenses/by/4.0/}{This work is licensed under a Creative Commons Attribution International 4.0 License.}
  \end{minipage}
  \vspace{5pt}
}
\makeatother

\begin{document}

\title[Attitudes and Expectations in Justice-Centered Data Structures \& Algorithms for Non-Majors]{``It Can Relate to Real Lives'': Attitudes and Expectations in Justice-Centered Data Structures \& Algorithms for Non-Majors}
% Student quotes for new possible titles:
    %% P1 %%
    % "The technology we use effects the world we live in"
    % "It can relate to real lives"
    % "grew in my ability to recognize the cultural and social significance of data structures and algorithms"
    % "view programming as an art that has many different effects on society"
    % "I have learned how to think in a more ethical way to use the technology."
    % "taught me additional new perspectives on programming and how to approach different forms of data"
    % "Teaching us about ethical considerations we have to make as coders gave me a new perspective"

    %% M1 %%
    % "I thought this course would have more "actual" coding"

\author{\an{Anna Batra}{Anonymous}}
\orcid{\an{0009-0007-5771-3270}{}}
\affiliation{%
  % \institution{\an{Computational Linguistics}{Institution}}
  \institution{\an{University of Washington}{Institution}}
  \city{\an{Seattle}{City}}
  \state{\an{WA}{State}}
  \country{\an{USA}{Country}}
}
\email{\an{batraa@uw.edu}{email}}

\author{\an{Iris Zhou}{Anonymous}}
\orcid{\an{0009-0003-4219-2768}{}}
\affiliation{%
  % \institution{\an{Department of Mathematics}{Institution}}
  \institution{\an{University of Washington}{Institution}}
  \city{\an{Seattle}{City}}
  \state{\an{WA}{State}}
  \country{\an{USA}{Country}}
}
\email{\an{irisz1@uw.edu}{email}}

\author{\an{Suh Young Choi}{Anonymous}}
\orcid{\an{0009-0006-2247-2254}{}}
\affiliation{%
  % \institution{\an{Department of Classics}{Institution}}
  \institution{\an{University of Washington}{Institution}}
  \city{\an{Seattle}{City}}
  \state{\an{WA}{State}}
  \country{\an{USA}{Country}}
}
\email{\an{atobdura@uw.edu}{email}}

\author{\an{Chongjiu Gao}{Anonymous}}
\orcid{\an{0009-0000-9547-4740}{}}
\affiliation{%
  % \institution{\an{Paul G. Allen School of Computer Science & Engineering}{Institution}}
  \institution{\an{University of Washington}{Institution}}
  \city{\an{Seattle}{City}}
  \state{\an{WA}{State}}
  \country{\an{USA}{Country}}
}
\email{\an{chongjiu@uw.edu}{email}}

\author{\an{Yanbing Xiao}{Anonymous}}
\orcid{\an{0009-0003-6335-5730}{}}
\affiliation{%
  % \institution{\an{Department of Electrical & Computer Engineering}{Institution}}
  \institution{\an{University of Washington}{Institution}}
  \city{\an{Seattle}{City}}
  \state{\an{WA}{State}}
  \country{\an{USA}{Country}}
}
\email{\an{ybx0525@uw.edu}{email}}

\author{\an{Sonia Fereidooni}{Anonymous}}
\orcid{\an{0009-0008-4522-8953}{}}
\affiliation{%
  % \institution{\an{Paul G. Allen School of Computer Science & Engineering}{Institution}}
  \institution{\an{University of Washington}{Institution}}
  \city{\an{Seattle}{City}}
  \state{\an{WA}{State}}
  \country{\an{USA}{Country}}
}
\email{\an{fereison@uw.edu}{email}}

\author{\an{Kevin Lin}{Anonymous}}
\orcid{\an{0000-0001-9946-3635}{}}
\affiliation{%
  % \institution{\an{Paul G. Allen School of Computer Science & Engineering}{Institution}}
  \institution{\an{University of Washington}{Institution}}
  \city{\an{Seattle}{City}}
  \state{\an{WA}{State}}
  \country{\an{USA}{Country}}
}
\email{\an{kevinl@cs.uw.edu}{email}}

\begin{abstract}
Prior work has argued for a more justice-centered approach to postsecondary computing education by emphasizing ethics, identity, and political vision. In this experience report, we examine how postsecondary students of diverse gender and racial identities experience a justice-centered Data Structures and Algorithms designed for undergraduate non-computer science majors. Through a quantitative and qualitative analysis of two quarters of student survey data collected at the start and end of each quarter, we report on student attitudes and expectations.

Across the class, we found a significant increase in the following attitudes: computing confidence and sense of belonging. While women, non-binary, and other students not identifying as men (WNB+) also increased in these areas, they still reported significantly lower confidence and sense of belonging than men at the end of the quarter. Black, Latinx, Middle Eastern and North African, Native American, and Pacific Islander (BLMNPI) students had no significant differences compared to white and Asian students.

We also analyzed end-of-quarter student self-reflections on their fulfillment of expectations prior to taking the course. While the majority of students reported a positive overall sentiment about the course and many students specifically appreciated the justice-centered approach, some desired more practice with program implementation and interview preparation. We discuss implications for practice and articulate a political vision for holding both appreciation for computing ethics and a desire for professional preparation together through iterative design.
\end{abstract}



\maketitle
\renewcommand{\shortauthors}{Anna Batra et al.}

\section{Introduction}

This experience report presents a study of the attitudes and expectations of postsecondary students enrolled in a justice-centered Data Structures and Algorithms (DSA) course designed for non-majors\footnote{\url{https://courses.cs.washington.edu/courses/cse373/23wi/}} at the University of Washington, Seattle campus, a large, research-intensive university in the United States. Our university is located in a growing high-tech area with many large tech companies and emerging startups that employ a majority white and Asian workforce. The University of Washington plays a key role in enabling this tech workforce by teaching computing skills to students from all majors across campus. At our university, there are two DSA courses: one only open to computer science majors, and one only open to non-computer science majors. Our paper focuses on the latter course. 

Given that justice-centered approaches have not been significantly studied in the postsecondary context, and because sociological phenomena like career funneling operate on larger scales beyond the scope of an individual course or even individual university, we wanted to focus on a more immediate unit of analysis: our students enrolling and taking computing courses today. The end goal of this study is to inspire future work to create, improve, and evaluate pedagogies for justice-centered approaches to higher computing education. We focus on these 2 research questions:

\begin{enumerate}[label=\textbf{RQ\arabic*}, leftmargin=*]
    \item How do students' computing confidence, perceptions of computing, and sense of belonging change (if at all) before and after taking a justice-centered DSA course?
    \item What are students' sentiments and experiences around their goals and expectations in a justice-centered DSA course?
\end{enumerate}

Prior work has explored the factors affecting confidence and sense of belonging for students, some contextualizing these factors in relation to students' gender and racial identities \cite{weird_but_okay, gender_and_confidence}. A study by \citeauthor{fears_confidence} explored differences in fears and confidence between students (primarily non-majors) taking two introductory programming courses: one more traditional course and one intentionally designed for non-majors. They found that students taking the course intentionally designed for non-majors, which included more creative programming assignments, experienced greater increases in confidence than students taking the more traditional course \cite{fears_confidence}. \citeauthor{i_can_do_that} surveyed and interviewed high school students taking introductory CS courses revealing that, although women experienced greater increases in self-efficacy than men, they still had lower efficacy than men overall at both the start and end of the term. Women initially had lower sense of belonging than men, but significantly increased in sense of belonging in the CS student community, with no significant differences at the end of the quarter \cite{i_can_do_that}. Another study by \citeauthor{BLNPI} found that women consistently had lower sense of belonging across three quarters of lower-level computing courses, but that Black, Latinx, Native American, and Pacific Islander (BLNPI) students did not have a lower or higher sense of belonging \cite{BLNPI}.

Although prior work has examined student expectations with respect to several dimensions of computing education with the relationship between students' computing experiences and their decisions to pursue a computer science major \cite{interest_in_CS_major}, to our knowledge there has not been prior work examining student expectations for DSA specifically. DSA might be one of the more challenging cases for implementing a justice-centered approach in the postsecondary computing curriculum because (1) it often follows multiple terms of introductory computing that might reinforce dominant narratives, and (2) it is often preceded by its reputation of being the course for preparing students for the ``technical interview.'' For these reasons, evaluating student goals and expectations might be able to surface a particularly nuanced view of the broader sociological forces shaping students' experiences in the course.

\section{Justice-Centered Approach}

At the heart of a justice-centered approach is an emphasis on students and the potential role that education can play in helping students understand the world around them, find their place within it, and empower them to effect the change that they wish to see. There is a growing body of work in culturally-relevant, culturally-responsive, and culturally-sustaining computing pedagogies in primary and secondary computing education that can help realize the goals of justice-centered computing education \cite{Vakil2018EthicsEducation, Vakil2020IveEducation, Ryoo2019PedagogyAll, Ryoo2020TakeSchools, Davis2021CulturallyFramework, Madkins2019CulturallyPractice}.

However, postsecondary computing education poses a particularly challenging context for this transformative approach to computing education. The dominant narrative for pursuing computing education emphasizes maximizing efficiency and business profit over ``do[ing] something good'' \cite{Vakil2018EthicsEducation, Vakil2020IveEducation, Ko2020ItEducation, Malazita2019InfrastructuresSubjects}. Recognizing the disciplinary values interpretation \cite{Vakil2020IveEducation} at work, more recent work has called on computing educators and researchers to address inequalities by drawing on diverse forms of computer science capital \cite{Kallia2021Re-ExaminingPerspective} and by addressing critical questions about the goals of computing education \cite{Washington2020WhenEnough, Lin2022CSNeed, Grosland2019ThroughPedagogy, Philip2018WhyWarfare, Philip2021TheoriesComputing, Vakil2019TheExplorations, Tissenbaum2021TheEducation, Ko2020ItEducation}.

Our particular course largely follows \emph{Critical Comparative Data Structures and Algorithms} (CCDSA), a justice-centered pedagogy for DSA that emphasizes a motif of ``critical comparison'' as a means of realizing ethics in the curriculum, identity in the learning environment, and political vision \cite{Lin2022CSNeed}.
\begin{description}[leftmargin=\parindent, itemsep=1ex]
    \item[Ethics via epistemological comparison] by teaching not only asymptotic analysis for data structure and algorithm implementations, but also affordance analysis \cite{Lin2021DoAnalysis} for ethical evaluation of data structure and algorithm interfaces.
    \item[Identity via cultural comparison] by drawing on core components for \emph{Culturally Responsive-Sustaining Computer Science Education} \cite{Davis2021CulturallyFramework}, such as providing equitable opportunities for creative and meaningful ways to learn and express learning.
    \item[Political vision via narrative comparison] by foregrounding diverse career paths that students can pursue after taking the course, refuting the narrative that DSA is (only) for fueling the elite tech sector.
\end{description}

Our course makes a novel contribution in how it bridges two key concepts that were left separate in CCDSA: analyzing implementations versus analyzing interfaces. In the dominant approach to DSA, students only analyze implementations for their runtime and use their analysis to create more efficient implementations. In CCDSA, students do all of the above but also learn to analyze interfaces for their affordances and evaluate the relationship between those affordances and real-world applications that utilize those interfaces. Our approach goes one step further to connect the two skills: to respond to affordance analysis, the software designer (the student) will need to design a new interface, which will then require new implementations that respond to asymptotic analysis.

This process then forms a nested loop: by iterating through both the outside loop (affordance analysis) and the inside loop (asymptotic analysis), our goal was for students to develop not only a greater appreciation for the role of ethics in the curriculum, but also a better understanding of their responsibility as software creators going beyond asymptotic analysis. When future interviewers ask one of our students to solve a coding challenge, we hoped they would be able to integrate both `inside the box' thinking (developing more efficient solutions) and `outside the box' thinking (surfacing unstated assumptions and designing new interfaces). In doing so, they would be able to not only preach ethics, but also put their ideas into practice too.

% Implementers should consider the impact of the product they are creating and can advocate for changes feasible within their work. 

% inform even the most mundane aspects of their future professional responsibilities. 

% - boring tasks with just aymptotic analysis; before it wasn't connected to the real world
% - they have more responsibility over the software they produce than they think. They can still do small things and advocate for it

To structure this process of \textbf{iterative design}, we drew on a technique for teaching critical software design skills called \emph{CIDER} \cite{Oleson2023TeachingTechnique}. In CIDER, students learn to address design biases and think beyond `average' users by
\begin{enumerate*}
    \item \textbf{C}ritiquing a real-world technology to identify a list of assumptions about users,
    \item \textbf{I}magining how one of the identified assumptions could lead to exclusion,
    \item (re)\textbf{D}esigning the technology to no longer rely on the assumption,
    \item \textbf{E}xpanding their knowledge by sharing their list of assumptions with others, and
    \item \textbf{R}epeating steps (2) and (3) to practice the work of addressing assumptions.
\end{enumerate*}
Iterative design has the potential to not only teach students a way to address design biases, but also to empower students to make decisions with an appreciation and more nuanced understanding of their capacity to change human conditions.

% New bit follows:
% Our implementation of CCDSA was premised on students having some familiarity with object-oriented programming in Java, so that we could use the Java programming language as context for learning, rather than an objective of learning. As such, we spent less time teaching students programming syntax and more time on technical and ethical details of various data structures and algorithms. 
Most coursework required students to submit self-recorded video explanations of their work. More specifically, we have:
\begin{itemize}
    \item conceptual assessments that evaluated student understanding of data structures and algorithms as they were introduced throughout the quarter;
    \item software projects where students implemented and analyzed solutions to real-world problems using the CCDSA approach;
    \item midterm interview questions where students were asked to come up with data structure and algorithm designs to address software problems;
    \item and a final portfolio where students explained how to implement a feature using graphs and apply iterative design to bring together all the aforementioned skills they learned in the course.
\end{itemize}

% Students completed two types of video-based assignments throughout the quarter: conceptual assessments that evaluated student understanding of data structures and algorithms as they were introduced throughout the quarter; and programming projects, where students focused on implementing and analyzing solutions to real-world problems using CCDSA. Video-based exams were more focused on design. In the midterm, students used Abstract Data Types (ADTs) and data structure design to create various approaches to solving prompts. On the final, students worked on designing knowledge graphs and graph-traversal algorithms to represent connections between related terms and to resolve search queries.
% through video submissions where they were expected to discuss the strengths and shortcomings of their problem-solving approaches; evaluate the effectiveness of their solutions using both asymptotic and affordance analysis; and reflect on their own experiences working on the projects.

\section{Study Population}

Our study population was comprised of students across two quarters: Autumn and Winter. Pre-quarter and post-quarter surveys were administered comprising questions about the following areas of interest: confidence in computing ability, perceived usefulness of computing, sense of belonging in CS, expectations and experiences, and students' identities. Out of the 426 total students still enrolled by the end of Autumn and Winter quarters, 334 students filled out both the pre-quarter and post-quarter surveys, giving us a response rate of 78.4\%. Only data from the 334 students who filled out both surveys were included in our analyses for our research questions. Both surveys were required for completion in the class, but participation in the study was optional.

Students reported their gender and racial identities using multiple option selection on the survey with an additional option to fill-in a response. Racial identities were grouped according to the 2020 census state redistricting issued by the US Department of Commerce in June 2021. Notable changes from the old census is that Middle Eastern and North African (MENA) are now in their own category
% is no longer considered as part of White/European and is their own category 
\cite{census_data} due to them not perceiving themselves as white, and others not perceiving them as white either \cite{MENA_not_white}.
Note that our course has both majority white and Asian student populations.

For quantitative analysis, we chose to group students with minority identities together when there were fewer than 30 observations. In terms of racial minorities, we grouped Black, Latinx, Middle Eastern and North African, Native American, and Pacific Islander (BLMNPI) students together. This practice was informed by \citeauthor{BLNPI}, which grouped Black, Latinx, Native American, and Pacific Islander (BLNPI) students together \cite{BLNPI}. Those of whom are mixed races are considered as part of every group they identified as we are using binary one-hot data encoding for our analysis in a one versus all approach. This explains why the racial identities do not all add up to 100\%. Similarly, we grouped women, non-binary, and other students not identifying as men (WNB+) because NB+ has less than 30 students in this group and may have more similar experiences to women rather than the majority gender, men.

\begin{table}[ht]
    \caption{Study participants' gender and racial identities}
    \centering
    \begin{tabular}{ll}
        \toprule
        % \multicolumn{2}{c}{Gender Identity} \\
        % \cmidrule(l){0-1}
        Men & WNB+ \\
        \midrule
        69\% & 31\% \\
        \bottomrule
    \end{tabular}
    \quad
    \begin{tabular}{lll}
        \toprule
        % \multicolumn{3}{c}{Racial Identity} \\
        % \cmidrule(l){0-2}
        White & Asian & BLMNPI \\
        \midrule
        14\% & 70\% & 10\% \\
        \bottomrule
    \end{tabular}
\end{table}

\section{RQ1 --- Student Attitudes}

The same 11 Likert items from the Attitudes Towards Computing Scale \cite{attitudes_scale} were posed at both the start and end of each academic quarter. The 11 statements fall under three categories, which we will refer to as the three computing attitudes:

\begin{enumerate}
    \item Confidence in computing and problem-solving abilities
    % statements focus on students' confidence in their ability to do advanced work or solve difficult problems in computing.
    \item Computing importance and perceived usefulness
    % statements focus on computing's influence on job opportunities and usefulness in life.
    \item Sense of belonging in the computing community
    % statements focus on whether students feel like they belong in CS as a field, take pride in their computing abilities, or consider themselves a scientist, technologist, engineer, or mathematician.
\end{enumerate}

Each student indicated their level of agreement with each statement on a Likert scale from 1 to 5, where 1 was ``Strongly Disagree'' and 5 was ``Strongly Agree.'' Some items in the scale are reverse-coded. During analysis, all items were standardized so that 1 represents a negative attitude and 5 represents a positive attitude.

Responses across each category of statements were averaged to create one mean composite score per category. This method of using mean composites within categories of questions was tested for reliability in the Attitudes Survey Validation Study \cite{attitudes_scale}. Within each category, we were interested in answering two questions:
\begin{enumerate}
    \item Did the class experience any significant changes in attitudes between the pre-quarter and post-quarter surveys?
    \item Did students of different gender or racial identities experience any significant differences in attitudes between the pre-quarter and post-quarter surveys?
\end{enumerate}
Separate statistical tests were used to answer each of the preceding questions. To determine whether any of the three computing attitudes increased, we used a 2-sample dependent t-test for each category of statements, pairing scores for each student from their pre-quarter response to their post-quarter response. To determine whether there were significant differences between responses across the three attitudes for different gender and racial groups, we used a 2-sample independent t-test. For each grouping, we tested to see if significant differences existed at the start and end of the quarter. We applied the Bonferroni correction for each grouping of students to account for multiple comparisons across the three categories of statements ($\alpha_{\textit{new}} = 0.05 / 3 \approx 0.017$).

\subsection{Results}

Table \ref{group-mean-table} shows the data. We summarize the results below with more information about the source of statistically significant differences.

\begin{table*}[ht]
    \caption{Pre-quarter and post-quarter mean composite scores for question categories where * indicates the ``yes'' score is significantly less than ``no'' score with $\alpha \approx 0.0167$. The pre and post averages for all students are displayed on the right.}
    \label{group-mean-table}
    \centering
    \begin{tabular}{rlllllllll}
        \toprule
        & & \multicolumn{2}{c}{WNB+} & \multicolumn{2}{c}{White} & \multicolumn{2}{c}{Asian} & \multicolumn{2}{c}{BLMNPI} \\
        \cmidrule(lr){3-4} \cmidrule(lr){5-6} \cmidrule(lr){7-8} \cmidrule(lr){9-10}
        & & Yes & No & Yes & No & Yes & No & Yes & No \\
    
        \midrule
        \multirow{2}{*}{Confidence} & pre & \textbf{3.15*} & 3.54 & 3.53 & 3.4 & \textbf{3.32*} & 3.64 & 3.41 & 3.42 \\
        & post & \textbf{3.41*} & 3.77 & 3.76 & 3.64 & 3.61 & 3.75 & 3.92 & 3.63 \\
    
        \midrule
        \multirow{2}{*}{Importance} & pre & 4.44 & 4.42 & 4.52 & 4.41 & 4.41 & 4.47 & 4.47 & 4.42 \\
        & post & 4.35 & 4.37 & 4.52 & 4.34 & 4.37 & 4.35 & 4.52 & 4.35 \\
    
        \midrule
        \multirow{2}{*}{Belonging} & pre & \textbf{3.39*} & 3.67 & 3.66 & 3.58 & 3.53 & 3.7 & 3.86 & 3.55 \\
        & post & \textbf{3.55*} & 3.8 & 3.88 & 3.7 & 3.72 & 3.73 & 3.94 & 3.7 \\
        \bottomrule
    \end{tabular}
    \quad
    \begin{tabular}{c}
        \toprule
        % Hack to simulate a cmidrule height
        \\[-2.35ex]
        \multirow{2}{*}{All} \\[0.65ex]
        \\
        \midrule
        3.42 \\
        3.65 \\
        \midrule
        4.43 \\
        4.36 \\
        \midrule
        3.58 \\
        3.72 \\
        \bottomrule
    \end{tabular}
\end{table*}

\subsubsection{Confidence in computing ability}

Across the class, students' confidence in their computing abilities significantly increased (p < .001) from the start of the quarter to the end of the quarter.
% Notably, three of the Likert statements within this category increased from a median of 3 (neutral) in the pre-quarter survey to a median of 4 (agree) in the post-quarter survey. The fourth reverse coded Likert item, ``I'm not good at computing'', invoked less change in responses, with a median of 2 (disagree) at both the start and the end of the quarter.
Across the groupings for gender and racial identities, significant differences in confidence were found in WNB+ and Asian students. WNB+ students reported significantly lower levels of confidence (p < .001) than men in both the pre and post-quarter surveys. Asian students also reported significantly lower levels of confidence at the start (p = .001), but not at the end of the quarter (p = .142) (Table \ref{group-mean-table}). Specifically, Asian students initially agreed less with the statement ``I think I could handle difficult computing problems,'' and agreed more with the statement ``I'm not good at computing.'' 
% Non-Asian students also decreased in agreement with the ``not good'' statement, contributing to the narrowed gap in scores.

No significant differences in confidence were found when grouping white students or BLMNPI students.

% However, BLMNPI students did notably increase by .51 in confidence, as compared to a .21 increase for white and Asian students.
% Both groups started with very similar mean composite scores 3.41 and 3.42 respectively [Table \ref{group-mean-table}].

\subsubsection{Computing importance and perceived relevance}

% Across the class, no significant change in perception of computing importance, usefulness, or relevance was found (p = 0.072).
No significant changes or differences were found in perception of computing importance across the class (p = .072) or with respect to gender and racial identities (p > .0167).
% The mean response decreased from 4.43 to 4.36, suggesting a small decrease in perceived computing importance and usefulness.

% Though these changes were not significant, Asian and WNB+ students decreased in perceived importance, while men and BLMNPI students increased in perceived importance.

\subsubsection{Sense of belonging in CS and in STEM}

Across the class, sense of belonging in CS and in STEM significantly increased (p < .001) from the start to the end of the quarter. Although median scores across the class did not change, students' responses to the statement, ``I feel like I 'belong' in computer science,'' were generally more neutral (median = 3) than for other statements about pride in computing abilities and self-consideration as STEM professional (median = 4). WNB+ students, additionally, reported significantly lower senses of belonging at both the start (p = .002) and end (p = .005) of the quarter. This was largely due to a more neutral response to the statement ``I feel like I belong in computer science'' in WNB+ students (mean = 3.05) than in men (mean = 3.54), at the start of the quarter.

No significant differences in sense of belonging were found when grouping white, Asian, or BLMNPI students.

\subsection{Discussion}

As this course is likely many students' first exposure to a collaborative, discussion-based computing course, the overall increase in confidence and sense of belonging is a welcome though unsurprising result. Encouraging collaboration and discussion may have led to a greater sense of community between students. Our results replicate prior studies that indicate a consistently lower sense of belonging in women and lack of difference between BLMNPI and non-BLMNPI students \cite{BLNPI, sax_belonging}.

% Although there was not a statistically significant difference between how much different groups of students increased in sense of belonging, we still observed small numerical differences. The class increased by .14, men increased by .13, and WNB+ students increased by .16. Other related works present potential justifications for these differences.

Findings from open-ended interviews conducted by \citeauthor{i_can_do_that} revealed that sense of belonging for female students was more related to collaboration with peers and female CS teachers \cite{i_can_do_that}. Our course's high percentage of 67\% WNB+ identifying teaching assistants might have contributed toward breaking down stereotypes about men being more suited toward technical roles, a perspective explored by Irani through qualitative interviews \cite{gender_and_confidence}. Implicit in the designation ``non-major'' is an out-group status label, which likely contributed to the baseline levels of belonging.
% Spitballing
% There are some results and numbers which we are not able to explain or infer any particular causation. For example, cultural differences and displays of confidence may impact a person's alignment with any particular attitude that we posed in our surveys.

Although prior work has explored differences in confidence and belonging with respect to gender, less work has examined differences in these attitudes for Asian or BLMNPI students. We found that Asian students had significantly lower computing confidence than non-Asian students at the start but not the end of the quarter (Table \ref{group-mean-table}). These results raise new questions considering the relative over-representation of Asians in our course compared to both the computing industry and broader demographics in the United States.

% Overall, students of different identities appear to have stronger reactions to some Likert statements than others. Future work through open-ended interviews could give more insight into how their background affects responses and why their feelings change.

\section{RQ2 --- Student Expectations}

In the post-quarter survey, we asked students to respond to a free-response question: \emph{Thinking back to the beginning of the quarter and your goals for taking this course, have you gained what you hoped to gain from this course? Was your experience in this class what you wanted or expected?} The open-ended nature of this question allowed students to discuss anything about the course, and students were also encouraged to write longer thoughts or reflections.

We inductively coded student responses to this question to understand their satisfaction with the course and how the course compared to their expectations. The first two passes were carried out by the first author to generate a preliminary codebook. Afterwards, three other authors, under supervision of the first author, each individually performed an additional pass adding onto the preliminary codebook. Preparation of the final results involved another review by the first author and merging of repeated codes.

\subsection{Results}
Our inductive coding produced an analysis of student sentiments (\ref{sentiment}) and an analysis of student experiences (\ref{experience}).

\subsubsection{Overall Sentiments}
\label{sentiment}

Each student response received exactly 1 of the following overall sentiment codes.

\begin{description}[leftmargin=\parindent, itemsep=1ex]
    \item[Positive] (86\% of class) for students who described an overall positive experience in the class, and indicates that they had a typical positive experience of satisfying their goal, such as ``Yes, the different algorithms.''
    \item[Negative] (13\% of class) for students who described an overall negative experience in the class, such as ``Harder than expected, intellij was kind of confusing for me.''
    \item[Inconclusive] (1.5\% of class) for students who described something that had an inconclusive sentiment, such as ``I slacked off a lot in this class so I didn't gain what I hoped to get out of this course.''
\end{description}

\begin{table}[ht]
    \centering
    \caption{Overall sentiments by identity}
    \label{overall-sentiments-by-identity}
    \begin{tabular}{rlllll}
        \toprule
        Code & Men & WNB+ & White & Asian & BLMNPI \\
        \midrule
        Positive & 84\% & 90\% & 87\% & 89\% & 84\% \\
        Negative & 14\% & 10\% & 11\% & 9\% & 13\% \\
        \bottomrule
    \end{tabular}
\end{table}

Table \ref{overall-sentiments-by-identity} shows the breakdown of positive and negative overall sentiments by identity. Due to inconclusive sentiments (not shown), not all columns sum to 100\%.

\subsubsection{Specific Experiences}
\label{experience}

Many students also received a code for their specific experiences in the class. A student response could receive 0, 1, or more than 1 of the following codes. Due to the open and optional nature of the prompts, not all students contributed specific experiences that corresponded to our research questions.

\begin{description}[leftmargin=\parindent, itemsep=1ex]
    \item[Appreciated Justice Approach (A-Justice)] for students who appreciated the justice-centered approach, such as ``This allow me to see how the code or technology we use effects the world we live in.'' \emph{More than 10 students received this code.}
    \item[Appreciated Change in Expectations (A-Change)] for students who initially expected more programming, but still appreciated what they learned in the course despite not matching their initial expectations, such as ``Discourse involves less coding assessment, than I thought, but I learned a much more useful thing.'' \emph{Fewer than 10 students received this code.}
    \item[Appreciated Programming and Interview Prep (A-Prep)] for students who appreciated improving their programming skills overall and preparation for software job interview questions, such as: ``LeetCode problems feel way easier now which is something I was hoping for.'' \emph{More than 10 students received this code.}
    \item[Desired Programming and Interview Prep (D-Prep)] for students who desired more implementation of data structures and algorithms concepts, usually toward the end goal of software jobs, such as ``it would be beneficial to have the professor share code with the class, and show us how to implement things.'' \emph{More than 10 students received this code.}
    \item[Desired Different Approach (D-Approach)] for students who desired a different course structure. \emph{Fewer than 10 students received this code.}
\end{description}

\subsection{Discussion} 
% D-Prep [M1] , A-Programming [P5], A-Interview [P6]
% START

\subsubsection{Programming and Interview Preparation}

We identified 3 types of desires under D-Prep: implementation practice, Java syntax, and technical interview preparation.

Students wanting more implementation felt that implementing data structures and algorithms, rather than learning through pseudo-code or conceptual problems, would best help them comprehend concepts. This desire for implementation may stem from the programming-heavy focus of our university's CS1 and CS2 courses. This is a focus that students may need to dis-acclimate from DSA in general, justice-centered or not, as portrayed by a student who ``expected a [computer science] course to be code focused,'' but realized that computing is ``more about the methods and analysis.''

Some students expected a continued focus on gaining familiarity with Java as a programming language. This expectation somewhat conflicted with our assumptions and intents; we assumed comfort with object-oriented programming in Java, and used Java as a context for learning rather than the focus of learning.

Additionally, some students also expected ``preparation and facilitated practice for coding interviews,'' since DSA concepts commonly appear in coding challenges and technical interviews. While this was not a focus of our course design, we found that students expressed valid concerns that they did not ``feel ready to pursue a professional career at many levels.''  Anecdotally, we note that our prior approach to DSA also failed to fully address these concerns.

%%%% STOP A-Prep and A-Interview %%%%
In contrast to the unmet expectations in D-Prep, students in A-Prep expressed satisfaction with the amount of implementation and Java practice. Some students reported that they could now ``solve LeetCode problems by [themselves] and know how to design an algorithm.'' Additionally, the self-recorded video-based explanation assessments helped them ``convey [their] thought process which\textellipsis{} help[ed] in preparing for technical interviews.'' It is noteworthy that some students felt more prepared for technical job interviews after taking our justice-centered computing course, as we don't spend time talking about how to prepare students for technical interviews that can be learned through geared practice of similar questions. Instead we focus on the on-the-job skills such as asymptotic and affordance analysis.

% CONTINUE
While incorporating more programming and interview-inspired assignments could be one way to meet student expectations in future quarters for D-Prep, there may be other ways to address student expectations. With the immense focus on the dominant software job narrative in the preceding CS1 and CS2 courses, our findings suggest that more effort could be directed to show how existing course practices prepare students for jobs of all kinds---in the tech industry and beyond.
%%%%%%%%%%%%%%%%%%%%%%%%%%%%%%%%%%%%%%%%%%%%%%%%%%%%%%%

%%% IMPLEMENTING DSA %%%

% A-Justice [P2] ----

% Start: 

\subsubsection{Change in Expectations and Different Approach}

A-Change represents students that expected this course to be purely technical and found the justice-centered concepts to be ``unique'' or ``unexpected,'' but ultimately appreciated that the inclusion of ethics gave them ``a new perspective'' on how ``it can relate to real lives.'' For example, one student stated ``my goal was to use this class to be prepared for technical interviews, which I did not accomplish because of the lack of coding, but I think I have since found an importance in understanding computing and the implications around it as I was taking the class.'' Importantly, the student does not seem to be expressing that the justice-centered approach caused their dissatisfaction or detracted from their other goals. Rather, their discovery of computing ethics seems to have changed their expectations. This suggests that there is room in the course design to hold both goals together: improve preparation for technical interviews and realize our political vision for students to view themselves as people whose actions can change the world. Another student recounts that they now ``view programming as an art that has many different effects on society.''

D-Approach included a student who had the inverse experience, where the inclusion of ethics was not a new perspective and thus ``did not challenge [their] thinking,'' since their major already focused on societal impacts. While 2 students mentioned this, our approach does indeed aim to introduce these ideas assuming no prior knowledge. Future offerings can consider presenting additional opportunities for deeper engagement with computing ethics by, for example, partnering with or highlighting current research efforts for socially-responsible computing at the institution. 

% Quote shows how student background affects reception of justice-centered approach. reference quote for above paragraph
% [D-Approach] I expected way more computational work. If anything, I think it ended up being way harder than I expected. Concepts like affordance analysis did not challenge my thinking because my focus is in the social studies side of Informatics, so I study ethics in information systems. This class tried to create a balance between ethics and tech, but it expected us to have a computational background (which I do not). Projects were terribly difficult because we never learned how to code them in lecture/section but were still expected to know how to implement them.

\subsubsection{Appreciated Justice-Centered Approach}

Additionally, other students in A-Justice expressed that aspects of assignment and course design presented avenues for personal growth or expressing their interests. One student noted a positive view of video-based explanation assessments. They described how, in previous programming courses, they ``had so many questions about those specs and the details/special cases'' and ``teachers were always like: `ignore it\textellipsis{} we don't require that for this assignment.''' In contrast, they felt that video submissions ``encouraged [them] to express [their] thoughts and assumptions'' and that it felt ``nice to see how those concerns for our programs are valued''---suggesting that their professional identity as a programmer could be better supported by assessment modalities that allowed for greater self-expression. They added that in-class discussion of affordance analysis, such as through lecture or iterative design activities, were ``very helpful and important for our study in CS.''

Another student noted that the open-ended, design-focused exams allowed them to ``connect what I learned in this course to my own personal interest in health informatics\textellipsis{} on the topic of hospital bed management.'' Some students noted that they learned a lot from this exam because they were ``able to review [their] `toolkit' of data structures and algorithms to answer different questions and find different approaches to them.'' In fact, one student expressed a desire for even more design questions as ``most of the projects already told us what structures to use'' and that more feedback on the design assignment may also help ``to better understand why [they] would use a specific data structure in a specific environment.''% Creative design or programming assignments give students avenues to explore their own ideas.

% "I knew from the first read through of the course home page that this class would be less about coding and more about real world scenarios. While I was disappointed at first\textellipsis{} Learning about affordance analysis and design techniques was a side of CSE that had never really been taught thoroughly before."

% Collaboration and teamwork were not only encouraged, but required for many of small-group discussions in larger programming assignments and lecture problem-solving activities. One student constructively criticized, ``sometimes the emphasis on teamwork outshadowed the focus on the cs concepts'', possibly referring to either the programming assignments or to lecture. They instead suggested ``really spending time having a class discussion on those [key concepts] and welcoming people to participate in that whole class discussion'', continuing that ``this would be such a wonderful experience to share with everyone in the class\textellipsis{} having the `Aha' moment together about how fundamental some concepts are.'' This suggests that the entire-class community can be a space for more generative and spontaneous interactions beyond those provided in the small-group discussions.

Collaboration and teamwork were not only encouraged, but required for many of small-group discussions in larger programming assignments and lecture problem-solving activities. In particular, one student constructively criticized, ``sometimes the emphasis on teamwork out-shadowed the focus on the CS concepts,'' possibly referring to either the programming assignments or to lecture. They instead suggested that ``really spending time having a class discussion on those [key concepts] and welcoming people to participate in that whole class discussion,'' would be a ``wonder experience to share with everyone'' and have the ```Aha' moment together about how fundamental some concepts are.'' This suggests that the entire-class community can be a space for more generative and spontaneous interactions beyond those provided in the small-group discussions.

One way this idea could be realized is by the structure of CCDSA. Another student commented, ``I loved that the concepts that we learned were taught in a way that always tied back to each other.'' The comparative data structures approach in the CCDSA pedagogy provides many opportunities for content to be connected to each other and revisited in different ways. These opportunities could be leveraged in class to spark discussion among students, rather than explained away in the individual pre-class preparation readings.

\section{Threats to Validity}

This experience report captures a very local snapshot of our offerings of a justice-centered DSA course for non-majors as we offered them over two quarters at our university. 

% We do not claim any causation for effects; even offering the same exact courses in the future may produce different results because of different students.

For the quantitative analysis of RQ1 in particular, note that our findings replicate effects found in dominant-approach courses. It is not impressive to us that taking another computing course would correlate with an increase in student confidence and sense of belonging given the circumstance and potential selection bias at work. 

% Furthermore, the computing importance and perceived relevance category may have exhibited a ceiling effect: average ratings were high for all students at the start and the end of the quarter. This null result is also unsurprising since DSA is a third or fourth course in computing at our university taken by non-majors---the majority of whom do not need it for graduation.

To provide a more fine-grained evaluation of the effects of instruction, it may be desirable to survey students more frequently than just at the start or end of the quarter. Another constraint of the survey format is the use of the Likert scale, which forced students to describe their attitudes and feelings within a 5-option discrete format without further justification. Additionally, even though students described their identities using multiple-option check-boxes with an optional fill-in box, their identities were analyzed within a categorical system of racial and gender classification---one that encodes mutually-exclusive binaries during its analysis. Nonetheless, the quantitative analysis offers a (limited) basis for future researchers to replicate.
% or scrutinize.

For the qualitative analysis of RQ2 in particular, one-on-one interviews or focus groups with students would provide a process to refine recommendations through dialogue rather than inference. These two offerings were among the first justice-centered offerings of any course offered by the computer science unit at our university, so most students likely had limited prior experience with critical computing concepts, which could increase discomfort and impact results.

\section{Conclusion}

This experience report makes three contributions to the field:
\begin{enumerate}
    \item an evaluation of postsecondary students in a justice-centered computing course, and recommendations for the future \cite{Lin2022CSNeed},
    \item a method for bridging the two conceptual halves of implementations and interfaces in CCDSA with CIDER \cite{Oleson2023TeachingTechnique},
    \item a justification, method, and set of results for evaluating postsecondary student expectations with inductive coding.
\end{enumerate}
Regarding student attitudes (RQ1), we replicated results from non-justice-centered courses in the literature about computing confidence and sense of belonging.
% and presented a possible ceiling effect for the importance of computing. 
Regarding student expectations (RQ2), we presented codes for overall sentiments and specific experiences of students after taking the course, as well as an in-depth discussion and reflection specific student experiences with potential implications for practice in both justice-centered DSA and dominant approaches to teaching DSA.

Future work should investigate justice-centered computing for in-majors, study the role and position of instructional staff in creating the conditions for justice-centered DSA, and lower barriers to adoption of critical computing pedagogies. Justice-centered computing has the potential to not only ``relate to real lives,'' but also empower students to provide better conditions for them too.

\bibliographystyle{ACM-Reference-Format}
\balance
\bibliography{ms}

\end{document}